\documentclass{ptephy_v1}

\preprintnumber{XXXX-XXXX} 

\usepackage{bm}
\usepackage{amsmath}
\usepackage{multirow}
\usepackage{color}

\newcommand{\lambdabar}{\lambda \kern -0.5em\raise 0.5ex \hbox{--}}

\begin{document}

\title{Avalanche Photon Cooling by Induced Compton Scattering: Higher-Order Kompaneets Equation}


\author{Shuta J. Tanaka$^{\ast}$, Katsuaki Asano and Toshio Terasawa}
\affil{Institute for Cosmic Ray Research, University of Tokyo, 5-1-5 Kashiwa-no-ha, Kashiwa City, Chiba, 277-8582, Japan \email{sjtanaka@icrr.u-tokyo.ac.jp}}


\begin{abstract}%
Induced Compton scattering (ICS) is an interaction between intense electro-magnetic radiations and plasmas, where ICS transfers the energy from photons to plasmas.
Although ICS is important for laser plasma interactions in laboratory experiments and for radio emission from pulsars propagating in pulsar wind plasmas, the detail of photon cooling process has not been understood.
The problem is that, when ICS dominates, evolution of photon spectra is described as a nonlinear convection equation, which makes photon spectra to be multi-valued.
Here, we propose a new approach to treat evolution of photon spectra affected by ICS.
Starting from the higher-order Kompaneets equation, we find a new equation that resolves the unphysical behavior of photon spectra.
In addition, we find the steady-state analytic solution, which is linearly stable.
We also successfully simulate the evolution of photon spectra without artificial viscosity.
We find that photons rapidly lose their energy by ICS with continuously forming solitary structures in frequency-space.
The solitary structures have the logarithmically same width characterized by an electron temperature.
The energy transfer from photons to plasma is more effective for broader spectrum of photons such as expected in astrophysical situations.
\end{abstract}

\subjectindex{xxxx, xxx}

\maketitle

\section{INTRODUCTION}\label{sec:Introduction}

Nonlinear interactions between strong electromagnetic waves and plasmas have been studied in the context of laser plasma physics and astrophysical phenomena.
Depending on intensities of radiations and conditions of plasmas, plenty of nonlinear interactions exist and are studied in different approaches.
For example, quantum electrodynamics predicts vacuum-polarization effects, where photons interact with virtual electron-positron pairs (c.f. Ref \cite{DiPiazza+12}).
Parametric processes in plasmas, such as induced scattering processes, filamentation instability and so on, have been studied in classical electrodynamics with the two-fluid description of plasmas (c.f. Ref. \cite{Kruer03}).
There are also studies of parametric processes in the semi-classical formulation, which is used in this paper (c.f. Refs. \cite{Dreicer64, Tsytovich70, Melrose08}).
Here, we focus on induced Compton scattering (ICS), which is a parametric instability between photons and each electron, rather than plasmon.
The condition for ICS to be dominant process is when both the central frequency and the spectral width of strong electromagnetic radiations are greater than the Langmuir plasma frequency \cite[][]{Galeev&Syunyaev73}.
ICS cools photons toward the Bose-Einstein condensation, while plasmas are heated by this process (c.f. Ref. \cite{Zel'dovich75}).

ICS has been studied in the context of laser plasma interaction both theoretically \cite[][]{Drake+74, Forslund+75, Lin&Dawson75} and experimentally \cite[][]{Decroisette+72, Drake+90, Leemans+91, Everett+95}.
Most of them studied the process with the two-fluid approximation, which accounts only for the linear regime of the instability.
On the other hand, ICS has also been studied for the scattering process around high intensity astrophysical sources with the semi-classical formulation, which can partly treat the nonlinear regime (e.g. Refs. \cite{Wilson&Rees78, Thompson+94, Lyubarsky08, Tanaka&Takahara13}). 
However, they constrained plasma conditions around some astrophysical objects considering only the scattering optical depth to ICS.
It is known that there is a difficulty to treat the nonlinear regime of ICS even with the semi-classical formulation as described in this paper.

In the semi-classical formulation, a relaxation process of isotropic photons interacting with a rarefied Maxwellian plasma by Compton scattering is described by the Kompaneets equation \cite[][]{Kompaneets57}.
The Kompaneets equation is the Fokker-Plank equation describing evolution of a photon spectrum including the ICS term, which comes from the Boson nature of photons and, which is quadratic in the intensity of the radiation.
Therefore, for high intensity radiations, this quadratic term plays a dominant role and the Kompaneets equation is reduced to the nonlinear convection equation (Eq. (\ref{eq:InducedOnly})). 
It is known that the nonlinear convection equation has the implicit solution, which will be multi-valued after a finite time starting from certain initial conditions.
Such a spectrum is unphysical and should be bent to be single-valued, i.e., there should be an appropriate physical process which we have ignored (e.g. Ref. \cite{Zel'dovich&Levich69}).

Different approaches have been adopted to resolve the unphysical behavior of a photon spectrum when ICS dominates.
Peyraud (1968) \cite{Peyraud68a, Peyraud68b, Peyraud68c} considered heuristically to add a dispersive term (the third derivative term) to the nonlinear convection equation and made the equation Korteweg-de Vries-type, i.e., soliton formation in frequency-space.
Reinish (1976) \cite{Reinish76a, Reinish76b} considered to recover the diffusive term (the second derivative term), which is originally included in the Kompaneets equation, and made the equation Burgers-type, which allows formation of a shock wave structure like hydrodynamics.
However, because the original diffusive term is linear in the intensity of radiations, this will not be applicable when the intensity of radiations becomes higher and higher.
On the other hand, Zel'dovich and colleagues \cite{Zel'dovich+72, Zel'dovich&Sunyaev72} adopted the integral form of the equation, i.e., the Boltzmann equation (Eq. (\ref{eq:BoltzmannEquation})).
They predicted formation of solitary structures in a radiation spectrum.
It is worth distinguishing which spectral behaviors appear in a given condition between photons and electrons.
There are also numerical studies of this problem by Montes (1977, 1979) \cite{Montes77, Montes79} and Coppi et al. (1993) \cite{Coppi+93}.
We compare their results with ours in Section \ref{sec:Discussion}.

In this paper, we propose a new approach to study evolution of a radiation spectrum when ICS dominates.
In Section \ref{sec:HigherOrderKompaneetsEquation}, we extend the Kompaneets equation to the higher-order terms.
This is a preparation to obtain a new equation and its derivation helps understanding what is the resolution of the problem in the nonlinear convection equation.
In Section \ref{sec:LargeOccupationNumber}, we find a new equation that describes evolution of a radiation spectrum for intense radiations.
We obtain the analytic solution in steady state and also discuss its linear stability.
In Section \ref{sec:NumericalSimulation}, we show some examples of numerical studies of the new equation.
We give interpretations of our results and comparison with past studies.
Section \ref{sec:Conclusions} is devoted to summary.

\section{Higher-Order Kompaneets Equation}\label{sec:HigherOrderKompaneetsEquation}

Here, we reconsider a kinetic equation that describes the evolution of the photon occupation number $n({\bm k})$ by Compton scattering.
Considering a rarefied plasma, we neglect emission and absorption of photons by plasmas, i.e., the photon number is conserved in this system (c.f. Ref. \cite{Zel'dovich75}).
Effects of the background magnetic field and polarizations of photons are also neglected.
Below, a wavenumber $k$ is in unit of the inverse of the electron Compton wavelength $\lambdabar_{\rm e} \equiv \hbar / m_{\rm e} c$ and a momentum $p$ of an electron is in unit of $m_{\rm e} c$.
We start from the Boltzmann equation for photons including Compton scattering by free electrons of a density $n_{\rm e}$.
We use a normalized momentum distribution of electrons $f({\bm p})$, i.e., $\int d^3 {\bm p} f({\bm p}) = 1$.
Assuming the spatial homogeneity of the system, we have
\begin{eqnarray}\label{eq:BoltzmannEquation}
	\frac{\partial n({\bm k})}{\partial t} 
	& = &
	c n_{\rm e} \int d^3 {\bm p} f({\bm p})
	\int \frac{d^3 {\bm k}_1}{k^2_1}
	\biggl[
	  D_1 n({\bm k}_1) (1 + n({\bm k  })) \left(\frac{k_1}{k} \right)^2 \frac{d \sigma}{d k   d {\bm \Omega  }} \nonumber \\
	  & - & 
	  D   n({\bm k}  ) (1 + n({\bm k_1}))                               \frac{d \sigma}{d k_1 d {\bm \Omega_1}} 
	\biggl],
\end{eqnarray}
where the terms $1 + n$ represent spontaneous and induced scattering terms, respectively.
$D = 1 - {\bm \beta} \cdot {\bm \Omega}$ and $D_1 = 1 - {\bm \beta} \cdot {\bm \Omega}_1$ represent the factor due to relative velocities between interacting photons and electrons, and $(k_1 / k)^2$ comes from the difference in phase space volumes between $d^3 {\bm k}_1$ and $d^3 {\bm k}$, where ${\bm \Omega} = {\bm k} / k$, ${\bm \Omega}_1 = {\bm k}_1 / k_1$, and $|{\bm \beta}| = (1 + p^{-2})^{-1/2}$, respectively.
The differential scattering cross section for Compton scattering (the Klein-Nishina cross section) from an initial state ${\bm k}_{\rm i}$ to a final state ${\bm k}_{\rm f}$ of a photon is (e.g. Ref. \cite{Pomraning73})
\begin{eqnarray}\label{eq:KleinNishinaObsFrame}
	\frac{d \sigma}{d k_{\rm f} d {\bm \Omega}_{\rm f}}
	& = &
	\frac{1}{\gamma D_{\rm f}} 
	\frac{3 \sigma^{}_{\rm T}}{16 \pi} \left( \frac{\tilde{k}_{\rm f}}{\tilde{k}_{\rm i}} \right)^2
	\delta \left(\tilde{k}_{\rm f} - \frac{\tilde{k}_{\rm i}}{1 + \tilde{k}_{\rm i} (1 - \tilde{\mu})} \right)
	\left(  \frac{\tilde{k}_{\rm f}}{\tilde{k}_{\rm i}} +  \frac{\tilde{k}_{\rm f}}{\tilde{k}_{\rm i}} - 1 + \tilde{\mu}^2 \right) \nonumber \\
	& = &
	\frac{3 \sigma^{}_{\rm T}}{16 \pi} \frac{1}{\gamma^2 D^2_{\rm i}} \left( \frac{k_{\rm f}}{k_{\rm i}} \right)^2
	\delta \left(k_{\rm f} - \frac{\gamma D_{\rm i} k_{\rm i}}{\gamma D_{\rm f} + k_{\rm i} (1 - \mu)} \right) \nonumber \\
	& \times &
	\left[ 1 + \left( 1 - \frac{1 - \mu}{\gamma^2 D_{\rm i} D_{\rm f}} \right)^2 + \frac{k_{\rm i} k_{\rm f} (1 - \mu)^2}{\gamma^2 D_{\rm i} D_{\rm f}} \right],
\end{eqnarray}
where quantities with $\tilde{}$ represent those in the electron-rest frame, $\gamma = (1 + p^2)^{1/2}$ is the Lorentz factor of an electron, $\mu = {\bm \Omega}_{\rm i} \cdot {\bm \Omega}_{\rm f}$ is the cosine of the angle between ${\bm k}_{\rm i}$ and ${\bm k}_{\rm f}$, $\delta$ is the Dirac's delta function, and $\sigma^{}_{\rm T}$ is the Thomson cross section.
We have used the following Lorentz transformation laws
\begin{eqnarray}\label{eq:LorentzTransformation}
	\tilde{k}_{\rm i, f} = \gamma D_{\rm i, f} k_{\rm i, f}, ~
	d \tilde{{\bm \Omega}}_{\rm f} =  \frac{d {\bm \Omega}_{\rm f}}{\gamma^2 D^2_{\rm f}}, ~
	\frac{d \sigma}{d \tilde{k}_{\rm f} d \tilde{{\bm \Omega}}_{\rm f}} = \gamma D_{\rm f} \frac{d \sigma}{d k_{\rm f} d {\bm \Omega}_{\rm f}},~
	1 - \tilde{\mu} = \frac{1 - \mu}{\gamma^2 D_{\rm i} D_{\rm f}},
\end{eqnarray}
and $(D_{\rm i} k_{\rm i})/(D_{\rm f} k_{\rm f}) + (D_{\rm f} k_{\rm f})/(D_{\rm i} k_{\rm i}) - 1 = 1 + k_{\rm i} k_{\rm f} (1 - \mu^2) / (\gamma^2 D_{\rm i} D_{\rm f})$ with the use of $\delta [k_{\rm f} - \gamma D_{\rm i} k_{\rm i} / (\gamma D_{\rm f} + k_{\rm i} (1-\mu))]$.

We execute the integral with respect to $k_1$ by using the $\delta$ function in Eq. (\ref{eq:KleinNishinaObsFrame}) and introduce notations $y = \Theta n_{\rm e} \sigma^{}_{\rm T} c t$, $\Theta \equiv k_{\rm B} T_{\rm pl} / m_{\rm e} c^2$, $k_{\pm} \equiv \gamma D k / (\gamma D_1 \mp k (1 - \mu))$, and $n({\bm k}_{\pm}) \equiv n(k_{\pm}, {\bm \Omega}_1)$, where $k_{\rm B}$ is the Boltzmann's constant and $T_{\rm pl}$ is an electron temperature.
Then, Eq. (\ref{eq:BoltzmannEquation}) is expressed as
\begin{eqnarray}\label{eq:BoltzmannEquation2}
	\frac{\partial n({\bm k})}{\partial y} 
	& = &
	\int \frac{4 \pi p^2 f(p) dp}{\Theta} \frac{d {\bm \Omega}_{\bm p}}{4 \pi} d {\bm \Omega}_1 [\Delta S + \Delta I],
\end{eqnarray}
where
\begin{eqnarray}\label{eq:Differences}
	\Delta S({\bm p}, {\bm k}, {\bm \Omega}_1)
	& \equiv &
	\frac{R_+ k^2_+ n({\bm k}_+) - R_- k^2_- n({\bm k})}{k^2 \gamma^2 D}, \label{eq:SpontaneousDifferences} \\
	\Delta I({\bm p}, {\bm k}, {\bm \Omega}_1)
	& \equiv &
	n({\bm k}) \frac{R_+ k^2_+ n({\bm k}_+) - R_- k^2_- n({\bm k}_-)}{k^2 \gamma^2 D}, \label{eq:InducedDifferences} \\
	R_{\pm}({\bm p}, {\bm k}, {\bm \Omega}_1)
	& \equiv &
	\frac{3}{16 \pi}
	\left[
		1 
		+ 
		\left( 1 - \frac{1 - \mu}{\gamma^2 D D_1} \right)^2 
		+ 
	\frac{k k_{\pm} (1 - \mu)^2}{\gamma^2 D D_1} \right]. \label{eq:DipoleTerm}
\end{eqnarray}
$\Delta S$ and $\Delta I$ correspond to contributions from spontaneous and induced scattering, respectively.
We find that net contributions to scattering are differences between an upper state ${\bm k}_+$ and the incident state ${\bm k}$ for spontaneous scattering, and between ${\bm k}_+$ and a lower state ${\bm k}_-$ for induced scattering.

Now, we assume that electrons and photons are non-relativistic, i.e., $p \ll 1$ and $k \ll 1$.
The lowest order in $p$ and $k$ gives Thomson scattering where $k_{\pm} \approx k$, $n({\bm k}_{\pm}) \approx n(k, {\bm \Omega}_1)$, and $R_{\pm} \approx 3 (1 + \mu^2) / 16 \pi$.
The result is
\begin{eqnarray}\label{eq:ZerothOrder}
	\Delta S_0
	& = &
	(1 + \mu^2) (n(k, {\bm \Omega}_1) - n(k, {\bm \Omega}))
	, \\
	\Delta I_0
	& = &
	(1 + \mu^2) n(k, {\bm \Omega}) (n(k, {\bm \Omega}_1) - n(k, {\bm \Omega}_1))
	= 0,
\end{eqnarray}
where $\Delta S_{(\xi)}$ and $\Delta I_{(\xi)}$ represent terms on an order $\xi$ for $\Delta S$ and $\Delta I$, respectively.
When the photon distribution is isotropic, we obtain $\Delta S_0 = 0$.
On the other hand, $\Delta I_0 = 0$ is always satisfied because ICS requires the frequency shift in essence.

On the first order with an assumption of isotropy, we obtain the well-known Kompaneets Equation.
The right-hand side of Eq. (\ref{eq:BoltzmannEquation2}) becomes
\begin{eqnarray}\label{eq:FirstOrder}
	& &
	\int \frac{4 \pi p^2 f(p) dp}{\Theta} 
	\int (\Delta S_{(k)} + \Delta I_{(k)} + \Delta S_{(p^2)})
	\frac{d {\bm \Omega}_{\bm p}}{4 \pi} d {\bm \Omega}_1 \nonumber \\
	& &
	= 
	\int \frac{4 \pi p^2 f(p) dp}{\Theta}
	\left[
		\frac{1}{k^2}
		\frac{\partial}{\partial k} k^4 \left(n + n^2 + \frac{p^2}{3} \frac{\partial n}{\partial k} \right)
	\right]
	= 
	\frac{1}{\Theta k^2}
	\frac{\partial}{\partial k} k^4 \left(n + n^2 + \Theta \frac{\partial n}{\partial k} \right),
\end{eqnarray}
where we have used $\int 4 \pi p^4 f(p) dp = 3 \Theta$, i.e., the Maxwell-Boltzmann distribution.
Though the result is simple, the derivation is fairly long.
We just note the expansion of $k_{\pm}$ on the first order in $k$ and $p^2$,
\begin{eqnarray}\label{eq:FirstOrderkpm}
	\frac{k_{\pm}}{k}
	& \approx &
	1 
	+
	p            (\alpha_1 - \alpha)
	+
	p^2 \alpha_1 (\alpha_1 - \alpha)
	\pm
	k (1 - \mu),
\end{eqnarray}
where $\alpha = {\bm \Omega}_{\bm p} \cdot {\bm \Omega}$ and $\alpha_1 = {\bm \Omega}_{\bm p} \cdot {\bm \Omega}_1$.
Odd orders in $p$ will vanish because of isotropy ($\int d {\bm \Omega}_1 d {\bm \Omega}_{\bm p} \alpha^{2m-1} \alpha^{2l-1}_1 = 0$ with natural numbers $m$ and $l$) and an order of $p^2$ corresponds to the first order of the plasma energy, i.e., the temperature $\Theta$.
Introducing normalized frequency $x = k / \Theta$, Eqs. (\ref{eq:BoltzmannEquation2}) and (\ref{eq:FirstOrder}) give the Kompaneets equation \cite[][]{Kompaneets57}
\begin{eqnarray}\label{eq:KompaneetsEquation}
	\frac{\partial n(x)}{\partial y}
	& = &
	\frac{1}{x^2} \frac{\partial}{\partial x}
	x^4
	\left(
		n(x)
		+
		n^2(x)
		+
		\frac{\partial n(x)}{\partial x}
	\right).
\end{eqnarray}
In addition, multiplying Equation (\ref{eq:KompaneetsEquation}) by $x^3$ and integrating over $d x$, we obtain evolution of the photon energy density ($\epsilon_{\rm ph} \equiv \int x^3 n(x) d x$) as (c.f. Ref. \cite{Levich&Sunyaev70})
\begin{eqnarray}\label{eq:PhotonEnergyEvolution}
	\frac{d \epsilon_{\rm ph}(y)}{d y}
	& = &
	- \int x^4 n(x) d x
	- \int x^4 n^2(x) d x
	+ 4 \epsilon_{\rm ph}.
\end{eqnarray}
The first term, which comes from $\Delta S_{(k)}$ of Eq. (\ref{eq:FirstOrder}), means energy loss of photons by spontaneous Compton scattering.
Corresponding induced term ($\Delta I_{(k)}$) is the second term, which also implies energy loss of photons.
On the other hand, the last term corresponding to $\Delta S_{(p^2)}$ expresses energy gain by inverse Compton scattering, even though this term comes from the elastic approximation (the Thomson limit, zeroth order in $k$).
It should be noted that there is no induced term associated with inverse Compton scattering, i.e., $\Delta I_{(p^2)} = 0$.
As is well-known, this Kompaneets equation is not enough to follow evolution of the photon spectrum for $n \gg1$.

Now, we proceed to the next order.
We have $\Delta S_{(k^2)}, \Delta S_{(k p^2)}$, and $\Delta S_{(p^4)}$ for spontaneous scattering but only a cross term $\Delta I_{(k p^2)}$ exists for induced scattering, because the numerator of Eq. (\ref{eq:InducedDifferences}) vanishes for even orders in each $k$ and $p$.
The expansion of $k_{\pm}$ on the second order in $k$ and $p^2$ ($\Theta$) is
\begin{eqnarray}\label{eq:SecondOrderkpm}
	\frac{k_{\pm}}{k}
	& \approx &
	1 
	+
	p            (\alpha_1 - \alpha)
	+
	p^2 \alpha_1 (\alpha_1 - \alpha)
	+
	p^3          (\alpha_1 - \alpha) \left( \alpha^2_1 - \frac{1}{2} \right)
	+
	p^4 \alpha_1 (\alpha_1 - \alpha)      ( \alpha^2_1 - 1                 ) \nonumber \\
	& \pm &
	k     (1 - \mu)
	+
	k^2   (1 - \mu)^2
	\pm
	k p   (1 - \mu)      (2 \alpha_1   -            \alpha)
	\pm
	k p^2 (1 - \mu) \left(3 \alpha^2_1 - 2 \alpha_1 \alpha - \frac{1}{2} \right).
\end{eqnarray}
Note that only the terms $\propto k$ have double sign on the right-hand side of Eq. (\ref{eq:SecondOrderkpm}) and this is the reason for $\Delta I_{(k^2)} = \Delta I_{(p^4)} = 0$, i.e., odd orders in $k$ is required for induced scattering.
The integrals are carried out as
\begin{eqnarray}\label{eq:Order_k^2}
	\int \frac{4 \pi p^2 f(p) dp}{\Theta} 
	\int \Delta S_{(k^2)} \frac{d {\bm \Omega}_{\bm p}}{4 \pi} d {\bm \Omega}_1 
	& = &
	\int \frac{4 \pi p^2 f(p) dp}{\Theta}
	\left(
		\frac{7}{10 k^2} \frac{\partial}{\partial k} k^6 \frac{\partial n}{\partial k}
	\right),
\end{eqnarray}
\begin{eqnarray}\label{eq:Order_p^4}
	\int \frac{4 \pi p^2 f(p) dp}{\Theta} 
	\int \Delta S_{(p^4)} \frac{d {\bm \Omega}_{\bm p}}{4 \pi} d {\bm \Omega}_1
	& = &
	\int \frac{4 \pi p^2 f(p) dp}{\Theta}
	\left(
		\frac{7 p^4}{150 k^2}
		\frac{\partial^2}{\partial k^2} k^6 \frac{\partial^2 n}{\partial k^2}
	\right),
\end{eqnarray}
and
\begin{eqnarray}\label{eq:Order_kp^2_Spontaneous}
	& &
	\int \frac{4 \pi p^2 f(p) dp}{\Theta} 
	\int \Delta S_{(k p^2)} \frac{d {\bm \Omega}_{\bm p}}{4 \pi} d {\bm \Omega}_1
	\nonumber \\
	& &
	=	
	\int \frac{4 \pi p^2 f(p) dp}{\Theta}
	\left[
		\frac{7 p^2}{30 k^2} \frac{\partial}{\partial k} 
		\left(
			k^6 \frac{\partial^2 n}{\partial k^2}
			+
			\frac{\partial}{\partial k} k^6 \frac{\partial n}{\partial k}
		\right)
		+
		\frac{5 p^2}{6 k^2} \frac{\partial}{\partial k} k^4 n
	\right],
\end{eqnarray}
for spontaneous scattering and as
\begin{eqnarray}\label{eq:Order_kp^2_Induced}
	& & 
	\int \frac{4 \pi p^2 f(p) dp}{\Theta} 
	\int \Delta I_{(k p^2)} \frac{d {\bm \Omega}_{\bm p}}{4 \pi} d {\bm \Omega}_1
	\nonumber \\
	& & 
	=
	\int \frac{4 \pi p^2 f(p) dp}{\Theta} 
	\left[
		\frac{p^2}{30 k^2} \frac{\partial}{\partial k} k^4
		\left(
			28 k^2 n \frac{\partial^2 n}{\partial k^2} 
			-
			14 k^2 \left(\frac{\partial n}{\partial k  } \right)^2
			+
			42 k  \frac{\partial n^2}{\partial k  } 
			+
			25 n^2
		\right)
	\right],
\end{eqnarray}
for induced scattering.
We also need the relativistic correction to the Maxwell-Boltzmann distribution $f(p) = e^{-\gamma/\Theta} / 4 \pi \Theta K_2(\Theta^{-1})$, where $K_2$ is the modified Bessel function of the second kind of order 2.
The moments of the distribution are written as $\int 4 \pi p^2 f(p) dp = 1$,
\begin{eqnarray}
	\int 4 \pi p^4 f(p) dp 
	& \approx & 
	3 \Theta + \frac{15}{2} \Theta^2, \label{eq:FermiDirac_p^4} \\
	\int 4 \pi p^6 f(p) dp 
	& \approx & 
	15 \Theta^2. \label{eq:FermiDirac_p^6}
\end{eqnarray}
This correction arises another second-order spontaneous term on the order of $\Delta S_{(p^4)}$ in addition to Eq. (\ref{eq:Order_p^4}) because we have 
\begin{eqnarray}\label{eq:SecondOrder_p^2}
	\int \frac{4 \pi p^2 f(p) dp}{\Theta} 
	\int \Delta S_{(p^2)} \frac{d {\bm \Omega}_{\bm p}}{4 \pi} d {\bm \Omega}_1
	& = &
	\int \frac{4 \pi p^2 f(p) dp}{\Theta}
	\left(
		\frac{p^2}{3 k^2}
		\frac{\partial}{\partial k} k^4 \frac{\partial n}{\partial k}
	\right) \nonumber \\
	& \approx &
	\left(
		3 \Theta
		+
		\frac{15}{2} \Theta^2
	\right)
	\frac{1}{3 \Theta k^2}
	\frac{\partial}{\partial k} k^4 \frac{\partial n}{\partial k},
\end{eqnarray}
where the second term of the last expression is on the order of $\Delta S_{(p^4)}$ while the first term is already appeared in Eq. (\ref{eq:KompaneetsEquation}).

Combining all of them, we obtain the higher-order Kompaneets equation that satisfies conservation of photon number, and the Bose-Einstein distribution is the equilibrium solution for this equation (c.f., the study for Eq. (\ref{eq:KompaneetsEquation}) is given by Caflisch and Levermore (1986) \cite{Caflisch&Levermore86}).
Note that the resultant equation is exactly the same as Eqs. (15) and (16) of Challinor and Lasenby (1998) \cite{Challinor&Lasenby98}, although they took a different approach to derive the equations.

\section{The Case for Large Occupation Number: Steady-State Solution and Its Stability}\label{sec:LargeOccupationNumber}

We consider the case for large occupation number $n \gg 1$.
When we start from the first order (Eq. (\ref{eq:KompaneetsEquation})), for $n \gg 1$, we obtain
\begin{eqnarray}\label{eq:InducedOnly}
	\frac{\partial N(x)}{\partial y}
	-
	2 N(x) 
	\frac{\partial}{\partial x} N(x)
	& = &
	0.
\end{eqnarray}
where $N(x) \equiv x^2 n(x)$.
This is the nonlinear convection equation, which leads to an unphysical multi-valued solution.
The situation is completely analogous to the shock formation in hydrodynamics so that the presence of a viscosity (the diffusion term in the right-hand side) can be the resolution (c.f. Ref. \cite{Zel'dovich&Raizer67}).
The viscous term arises when we extend the ideal fluid (the Euler equations) to the viscous fluid (the Navier-Stokes equations) (c.f. Ref. \cite{Landau&Lifshitz59}). 
On the other hand, in this case of ICS, we will see that the physical resolution is not a diffusion term. 

Now, we start from the higher-order Kompaneets equation and find a new equation that resolves the difficulty of Eq. (\ref{eq:InducedOnly}).
Because we expand Eq. (\ref{eq:BoltzmannEquation}) assuming $k \ll 1$ and $\Theta \ll 1$ ($p \ll 1$), we neglect the second-order spontaneous terms (Eqs. (\ref{eq:Order_k^2}) $-$ (\ref{eq:Order_kp^2_Spontaneous}) and (\ref{eq:SecondOrder_p^2})) compared with the first-order spontaneous terms and we obtain
\begin{eqnarray}\label{eq:ExtendedKompaneetsEquation0}
	\frac{\partial n}{\partial y} 
	\approx
	\frac{1}{x^2} \frac{\partial}{\partial x} x^4
	\biggl[n + \frac{\partial n}{\partial x} + \left(1 - \frac{5}{2} \Theta\right) n^2
	+
	\frac{7}{5} \Theta
	\left(
		2 x^2 n   \frac{\partial^2 n}{\partial x^2} 
		-
		x^2 \left(\frac{\partial   n}{\partial x  } \right)^2
		+
		3 x       \frac{\partial n^2}{\partial x  } 
	\right) \biggl].
\end{eqnarray}
Efficiencies of each term on the right-hand side of Eq. (\ref{eq:ExtendedKompaneetsEquation0}) are characterized by scattering optical depths $\tau$.
In order of magnitude calculation, dividing each term by $n / y$, the right-hand side of Eq. (\ref{eq:ExtendedKompaneetsEquation0}) includes four components whose scattering optical depths are described as $\tau^{}_{\rm spon} = x y$ (the first term: spontaneous Compton), $\tau^{}_{\rm inv} = y$ (the second term: inverse Compton), $\tau^{}_{\rm ind} = x y n$ (the third term: induced Compton) and $\tau^{}_{\rm sec} = x y n \Theta \ll \tau^{}_{\rm ind}$ (the other terms proportional to $\Theta$: second-order induced Compton), respectively.
ICS becomes dominant process if $\tau^{}_{\rm ind} \gg \tau^{}_{\rm spon}, \tau^{}_{\rm inv}$, i.e., $n \times \min(1, x) \gg 1$.
If we consider only the $\tau^{}_{\rm ind}$ term, the mathematically troublesome Eq. (\ref{eq:InducedOnly}) is derived.
On the other hand, when the $\tau^{}_{\rm sec}$ terms are neglected, we obtain the usual Kompaneets equation Eq. (\ref{eq:KompaneetsEquation}).
For the case of $\tau^{}_{\rm sec} / \max(\tau^{}_{\rm spon}, \tau^{}_{\rm inv}) = n \Theta \times \min(1, x) \gg 1$, the first-order spontaneous terms are neglected compared with the second-order induced terms even though the $\tau^{}_{\rm sec}$ terms are higher-order in $k$ or $\Theta$.
This is because the spontaneous and induced terms have different dependence on $n$ so that this does not mean an importance of the third and further higher-order terms in $k$ or $\Theta$.
	
The large occupation number satisfying the condition $n \Theta \times \min(1, x) = \min(n \Theta, n k) \gg 1$ is expected in some situations, for example, $n \sim 10^{42}$ (brightness temperature $T_{\rm b}(\nu) \equiv h \nu n(\nu) / k_{\rm B} \sim 10^{41} {\rm K}$ at frequency $\nu \sim 10 {\rm GHz}$) is implied from an observation of the Crab pulsar in astrophysics \cite[][]{Hankins&Eilek07} and $n \sim 10^{22}$ (petawatt laser with $\sim$ 3.3 nm spectral width at wavelength $\sim 10.5 \mu$m) is obtained from some current laser facilities (e.g. Ref. \cite{Kawanaka+08}).
In this case, we find that the natural extension of Eq. (\ref{eq:InducedOnly}) is written as
\begin{eqnarray}\label{eq:ExtendedKompaneetsEquation}
	\frac{\partial N}{\partial y}
	-
	2 N
	\frac{\partial}{\partial x} N
	=
	\frac{14}{5} \Theta 
	\left(
		x N
		\frac{\partial^3}{\partial x^3} (x N)
		-
		\frac{17}{14} N
		\frac{\partial  }{\partial x} N
	\right).
\end{eqnarray}
Note that Eq. (\ref{eq:ExtendedKompaneetsEquation}) exactly corresponds to the isotropic case of Eq. (\ref{eq:BoltzmannEquation2}) with $\Delta S = 0$ and $\Delta I = \Delta I_{(k)} + \Delta I_{(k p^2)}$.
In contrast to Eq. (\ref{eq:InducedOnly}), the plasma temperature $\Theta$ appears explicitly on the right-hand side that corresponds to $\Delta I_{(k p^2)}$ (Eq. (\ref{eq:Order_kp^2_Induced})).
Because the right-hand side is the higher-order correction term, the velocity of photon flux (in frequency-space) is basically determined by the second term of the left-hand side, i.e., order $N(x)$ to the negative direction of $x$.
The first term of the right-hand side is important to resolve the problem described in Section \ref{sec:Introduction}, while the second term slightly modifies the second term of the left-hand side.
Eq. (\ref{eq:ExtendedKompaneetsEquation}) is similar to the Korteweg-de Vries equation, but the coefficient of the dispersive term is also non-linear (proportional to $N$).
This is different from the study by Peyraud (1968) \cite{Peyraud68a, Peyraud68b, Peyraud68c}.

\begin{figure}[t]
\begin{center}
\includegraphics[scale=0.70]{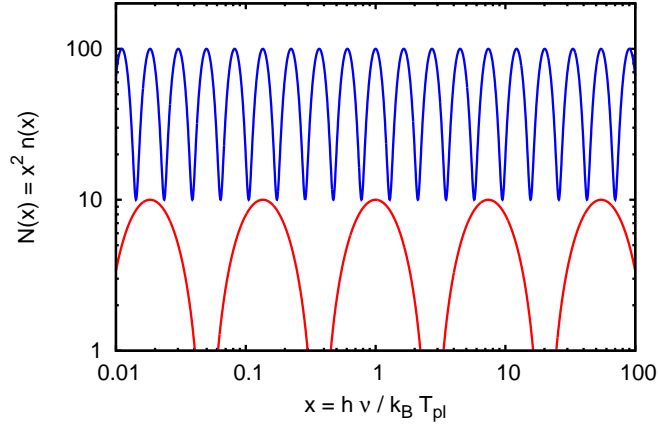}
\end{center}
\caption{
	Plots of Eq. (\ref{eq:SteadyStateSolution}) in logarithmic scale for different parameters: $(k_{\Theta}, A, B, \phi) = (\pi, 5, 5, 0)$ (red line) and $(4 \pi, 45, 55, 0)$ (blue line).
	The plasma temperature of the blue line is colder than that of the red one.
}
\label{fig:SteadyStateSolution}
\end{figure}

Before proceeding numerical calculations, let us find the steady-state solution.
The exact analytic solution of Eq. (\ref{eq:ExtendedKompaneetsEquation}) is obtained in steady state $\overline{N}(x)$, i.e.,
\begin{eqnarray}\label{eq:SteadyState}
	\frac{\partial \overline{N}}{\partial y}
	& = &
	\frac{14}{5} \Theta \overline{N}
	\left(
		x \frac{\partial^3}{\partial x^3} (x \overline{N})
		+
		(k^2_{\Theta} + 1) \frac{\partial \overline{N}}{\partial x}
	\right) \nonumber \\
	& = &
	\frac{\partial}{\partial x}
	\left[
		\frac{7}{5} \Theta
		\overline{N}^{\frac{3}{2}} 
		\left(
			4 x^2 \frac{\partial^2}{\partial x^2}
			+
			4 x \frac{\partial}{\partial x}
			+
			k^2_{\Theta}
			\right) \overline{N}^{\frac{1}{2}}
	\right]
	=
	0,
\end{eqnarray}
where we introduce the Doppler wavenumber $k_{\Theta} \equiv (5/(7 \Theta) -31/14)^{1/2} \gg 1$ for $\Theta \ll 1$.
We obtain a non-trivial solution ($\overline{N} \ne 0$) from the first line of Eq. (\ref{eq:SteadyState}) as 
\begin{eqnarray}\label{eq:SteadyStateSolution}
	\overline{N}(x)
	=
	A \cos(k_{\Theta} \ln x + \phi) + B,
\end{eqnarray}
where an amplitude $A (\ge 0)$, a DC component $B$ and a phase $\phi$ are constants of integration.
Eq. (\ref{eq:SteadyStateSolution}) shows no characteristic photon energy $x$, because the first-order spontaneous terms are neglected in Eq. (\ref{eq:ExtendedKompaneetsEquation}), where only $\tau^{}_{\rm inv}$ has different dependence on $x$ from $\tau^{}_{\rm spon}, \tau^{}_{\rm ind}$ and $\tau^{}_{\rm sec}$.
This steady-state solution $\overline{N}(x)$ includes the solution of Eq. (\ref{eq:InducedOnly}), as $A = 0$, i.e., the first term of Eq. (\ref{eq:SteadyStateSolution}) is the important contribution from $\Delta I_{(k p^2)}$.
We require $B \ge A$ for $\overline{N}(x) > 0$.
Inside the derivative of the second line of Eq. (\ref{eq:SteadyState}) is written as a constant $(7 \Theta / 5) k^2_{\Theta} (B^2 - A^2) \ge 0$, i.e., photon flux in frequency-space is negative direction of $x$.
The photon flux is zero for $A = B$, i.e., contribution from $\Delta I_{(k p^2)}$ compensates $\Delta I_{(k)}$ term in this case and plays as a kind of heating, while $\Delta I_{(k)}$ term plays a role of cooling (see Eq. (\ref{eq:PhotonEnergyEvolution})).
We require proper boundary conditions at finite frequencies to determine the constants of integration, because the number density $\int \overline{N}(x) dx$ and the energy density $\int x \overline{N}(x) dx$ of photons diverse for either $x \rightarrow 0$ or $x \rightarrow \infty$.
Since the photon flux is the negative direction, a photon injection at a large $x$ and absorption at a small $x$ are expected to achieve the steady-state (c.f., a similar study for Eq. (\ref{eq:KompaneetsEquation}) is given by Dubinov (2009) \cite{Dubinov09}).
We plot Eq. (\ref{eq:SteadyStateSolution}) in Fig. \ref{fig:SteadyStateSolution}, for example.
One of important features of Eq. (\ref{eq:SteadyStateSolution}) is that the Doppler wavenumber $k_{\Theta} \approx \Theta^{-1/2}$ associates with logarithmic scale in frequency $x$.
This arises from the third-derivative term in Eq. (\ref{eq:ExtendedKompaneetsEquation}) and is interpreted as the Doppler width $\Delta \nu \sim \Theta^{1/2} \nu$ by induced-inverse Compton process ($\Delta I_{(k p^2)}$) already discussed by Zel'dovich and Sunyaev (1972) \cite{Zel'dovich&Sunyaev72} in the integral form.
Note that this Doppler width is a different process from the first-order spontaneous inverse Compton scattering $\Delta S_{(p^2)}$.
For the later convenience, we define the Doppler wavelength $\lambda^{}_{\Theta} \equiv 2 \pi / k_{\Theta}$.
For every integer $m$, the peaks of $\overline{N}(x)$ appear at $x = \Phi e^{m \lambda^{}_{\Theta}}$ with a constant $\Phi = e^{- \lambda^{}_{\Theta} \phi / 2 \pi}$.

Finally, we analyze linear stability for the steady-state solution $\overline{N}(x)$.
We substitute $N(x, y) = \overline{N}(x) + N_1(x, y)$ ($N_1 /\overline{N} \ll 1$) into Eq. (\ref{eq:ExtendedKompaneetsEquation}) and linearize the equation,
\begin{eqnarray}\label{eq:Linearised}
	\frac{\partial N_1}{\partial y}
	& = &
	\frac{14}{5} \Theta \overline{N}
	\left(
		x \frac{\partial^3}{\partial x^3} (x N_1) 
		+
		(k^2_{\Theta} + 1)
		\frac{\partial N_1}{\partial x}
	\right).
\end{eqnarray}
Considering a Fourier component of $N_1(x, y) \propto e^{i(\omega y - \kappa x)}$, we obtain the dispersion relation
\begin{eqnarray}\label{eq:DispersionRelation}
	\omega 
	& = &
	\frac{14}{5} \Theta \overline{N}
	\left(
		i x \kappa^2
		-
		\kappa^3 x^2
		-
		(\kappa^2_{\Theta} + 1) \kappa
	\right).
\end{eqnarray}
Because ${\rm Im}(\omega) > 0$, the steady-state solution Eq. (\ref{eq:SteadyStateSolution}) is linearly stable.

\section{Numerical Simulation}\label{sec:NumericalSimulation}

Here, we numerically study the initial evolution of photon spectra by solving Eq. (\ref{eq:ExtendedKompaneetsEquation}) with $n \gg 1$.
We pay particular attention to evaluate the third-derivative in Eq. (\ref{eq:ExtendedKompaneetsEquation}) precisely in order to see developments of solitary structures quoted by Zel'dovich and colleagues \cite{Zel'dovich&Sunyaev72, Zel'dovich+72}.

\subsection{Set up}\label{sec:Setup}

Eq. (\ref{eq:ExtendedKompaneetsEquation}) is solved by fourth-order in $x$ derivative (five-point stencil) and fourth-order in $y$ derivative (fourth-order Runge-Kutta).
We use $x$ grid points spaced linearly from 0.1 to 10 and put fixed boundary conditions at both sides of $x$ boundaries.
Initial spectra are assumed to be a Gaussian spectrum,
\begin{eqnarray}\label{eq:Gaussian}
	N_0(x)
	& = &
	N_{\rm init}
	\exp\left( - \frac{(x - x_{\rm init})^2}{2 \sigma^2_{\rm init}} \right),
\end{eqnarray}
including three parameters, a normalization $N_{\rm init}$, a mean $x_{\rm init}$ and a half width $\sigma_{\rm init}$ of a Gaussian.
The ratio of the initial spectral width $\sigma_{\rm init}$ to the Doppler wavelength $\lambda^{}_{\Theta} \equiv 2 \pi / k_{\Theta}$, which is a parameter associated with Eq. (\ref{eq:ExtendedKompaneetsEquation}), characterizes the spectral evolution.
While sufficiently broad spectra ($\sigma_{\rm init} \gg \lambda^{}_{\Theta}$) are expected in astrophysical situations, spectral widths of laser beams in laboratories are not always broad.
We choose initial conditions that satisfy $n_{\rm init} \Theta \times \min(1, x_{\rm init}) \gg 1$ ($x^2_{\rm init} n_{\rm init} \equiv N_{\rm init}$).
Since the optical depth to ICS is written as $\tau_{\rm ind} = y N / x$, we measure time $y$ with $y_0 \equiv x_{\rm init} / N_{\rm init}$ and an adequate time-step is searched to resolve spectral evolution for each calculation.

Eq. (\ref{eq:ExtendedKompaneetsEquation}) is approximated form of Eq. (\ref{eq:ExtendedKompaneetsEquation0}) for $n \Theta \times \min(1, x) \gg 1$.
We checked that both of the equations give almost the same initial evolution until $y \lesssim 0.4 y_0$ for the same parameter sets, satisfying $n \Theta \times \min(1, x) \sim 10$ and $10^2$ at the peak of a Gaussian $x = x_{\rm init}$.
For $0.4 y_0 \lesssim y < y_0$, numerically unstable structures start to develop at the portion of $N(x, y) \ll N_{\rm init}$ in our current scheme.
This does not allow us to pursue the spectral evolution for different $\sigma_{\rm init}$ and $\lambda^{}_{\rm \Theta}$ beyond $y \sim 0.4 y_0$.
In what follows, we show numerical results of Eq. (\ref{eq:ExtendedKompaneetsEquation}) for $y \le 0.4 y_0$.
We fixed $(N_{\rm init}, x_{\rm init}) = (10^7, 1)$, since different sets of $(N_{\rm init}, x_{\rm init})$ does not change results as long as we measure $y$ with $y_0$.
For remaining two parameters, we study the case that both $\sigma_{\rm init}$ and $\lambda^{}_{\Theta}$ are an order of $10^{-1}$ ($\lambda^{}_{\Theta} = 0.1$ corresponds to an electron temperature of $\sim 10^2$ eV).

Eq. (\ref{eq:ExtendedKompaneetsEquation}) has two invariants of motion, the photon number density $\int N(x, y) dx$ and a quantity $\int \ln N(x, y) dx$ \cite[][]{Zel'dovich&Sunyaev72}.
Conservation of photon number is immediately found when we rewrite $x$ derivatives in Eq. (\ref{eq:ExtendedKompaneetsEquation}) to the second line expression of Eq. (\ref{eq:SteadyState}) and integrate both sides the equation over $d x$.
We obtain conservation of the quantity $\int \ln N(x, y) dx$ from dividing both sides of Eq. (\ref{eq:ExtendedKompaneetsEquation}) by $N(x)$ and integrating over $d x$.
For all the results of our calculation below, we checked that these quantities were conserved for $y \le 0.4 y_0$.

\subsection{Results}\label{sec:Results}

%
\begin{figure}[t]
\begin{minipage}{0.5\hsize}
\begin{center}
\includegraphics[scale=0.62]{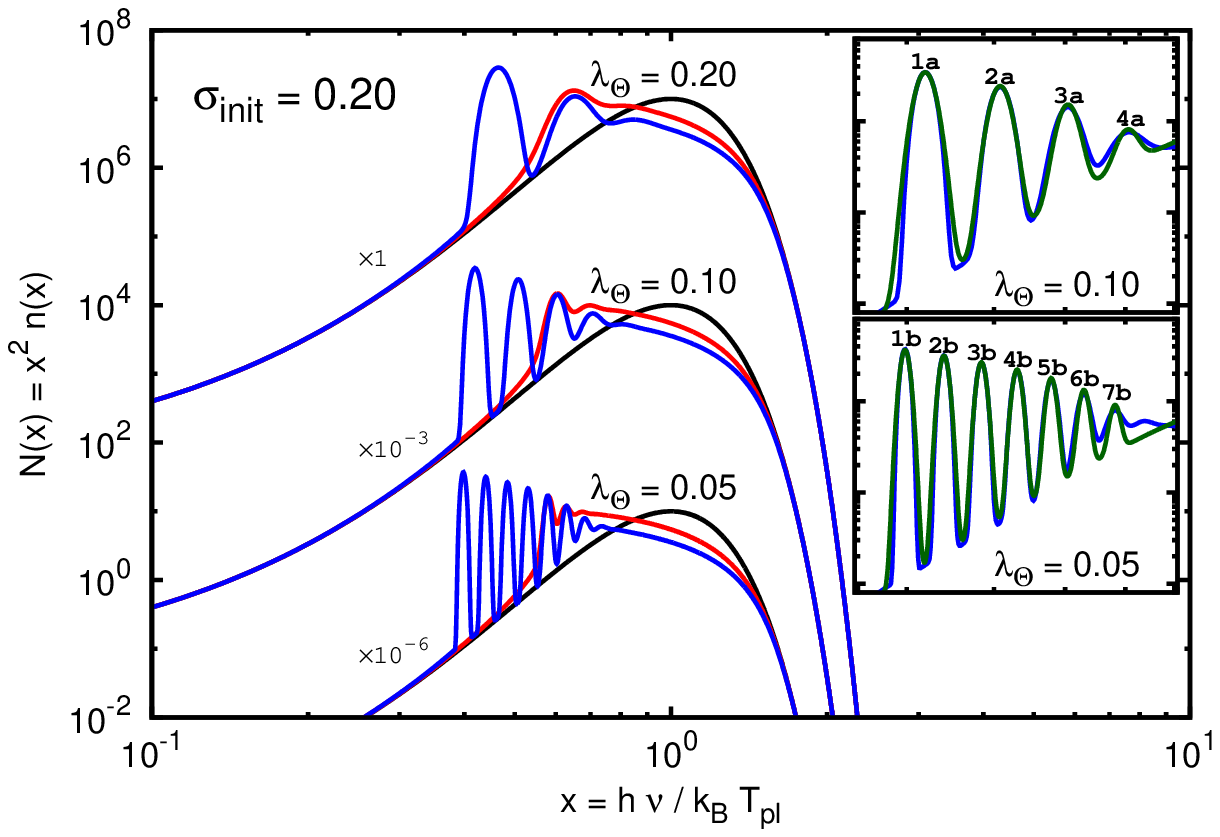}
\end{center}
\end{minipage}
\begin{minipage}{0.5\hsize}
\begin{center}
\includegraphics[scale=0.62]{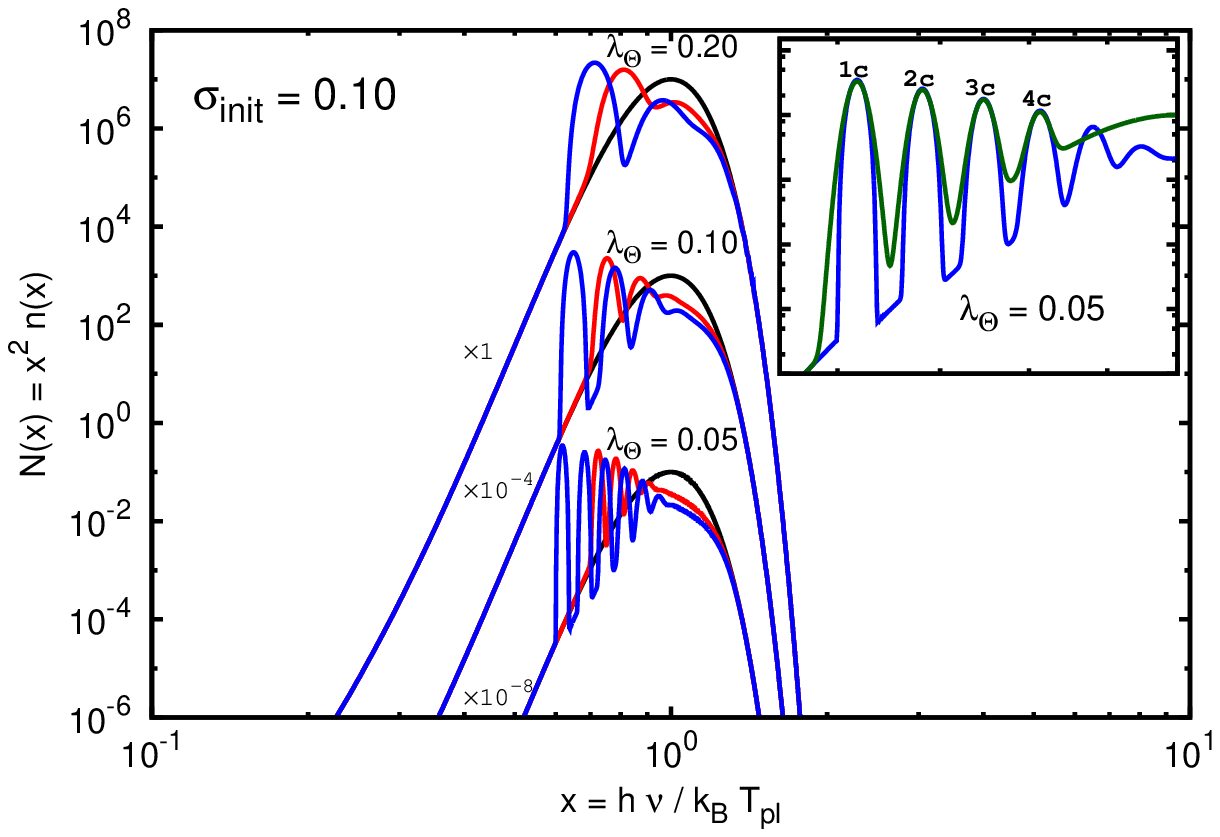}
\end{center}
\end{minipage}
\caption{
	Each panel shows results of numerical calculation of Eq. (\ref{eq:ExtendedKompaneetsEquation}) for three different Doppler width $\lambda^{}_{\Theta} \equiv 2 \pi / k_{\Theta} = 0.20$ (uppermost lines), $0.10$ (middle lines), and $0.05$ (lowermost lines).
	The initial spectral widths are different between left ($\sigma_{\rm init} = 0.2$) and right panels ($\sigma_{\rm init} = 0.1$), while $(N_{\rm init}, x_{\rm init}) = (10^7, 1)$ is common (see Eq. (\ref{eq:Gaussian})), i.e., $y_0 = 10^{-7}$ for all cases.   
	Black, red and blue lines correspond to $N(x, 0)$, $N(x, 0.2 y_0)$ and $N(x, 0.4 y_0)$, respectively.
	Insets are zoom of solitary structures appeared in the middle line ($\lambda_{\Theta} = 0.10$) and the lowermost line ($\lambda_{\Theta} = 0.05$) on the left panel and in the lowermost line ($\lambda_{\Theta} = 0.05$) on the right panel when $y = 0.4 y_0$.
	Blue lines in the insets are the result of calculation $N(x, 0.4 y_0)$ and green lines are Eq. (\ref{eq:Fitting}) with parameters tabulated in Table \ref{tbl:Fitting}.
	Numbering of solitary structures corresponds to the number $i$ on Table \ref{tbl:Fitting}.
}
\label{fig:Evolution}
\end{figure}
\begin{table}[!t]
\begin{minipage}{1.0\hsize}
\caption{
	Fitted values of $n_i$ and $m_i$ in Eq. (\ref{eq:Fitting}).
	Number $i$ of solitary structures is found in the insets in Fig. \ref{fig:Evolution}.
	Corresponding $\sigma_{\rm init}$ and $\lambda^{}_{\Theta}$ are also tabulated.
	$\Delta m_i$ is separation between the neighboring peaks of the $i$-th solitary structure, $\Delta m_{\rm 1a} = m_{\rm 2a} - m_{\rm 1a}$ for example.
}
\label{tbl:Fitting}
\begin{center}
\begin{tabular}{cccccc}
\hline
$i$& $\sigma_{\rm init}$ & $4 w = \lambda^{}_{\Theta}$ & $n_i$              & $m_i$  & $\Delta m_i$ \\
\hline
1a & \multirow{4}{*}{0.20} & \multirow{4}{*}{0.10} & $3.42 \times 10^7$ & -0.869 & 0.192  \\
2a &                       &                       & $2.38 \times 10^7$ & -0.677 & 0.173  \\
3a &                       &                       & $1.38 \times 10^7$ & -0.504 & 0.153  \\
4a &                       &                       & $4.78 \times 10^6$ & -0.351 & $-$    \\
\hline
1b & \multirow{7}{*}{0.20} & \multirow{7}{*}{0.05} & $3.64 \times 10^7$ & -0.921 & 0.0996 \\
2b &                       &                       & $3.14 \times 10^7$ & -0.821 & 0.0960 \\
3b &                       &                       & $2.64 \times 10^7$ & -0.725 & 0.0915 \\
4b &                       &                       & $2.16 \times 10^7$ & -0.634 & 0.0873 \\
5b &                       &                       & $1.68 \times 10^7$ & -0.547 & 0.0838 \\
6b &                       &                       & $1.15 \times 10^7$ & -0.463 & 0.0795 \\
7b &                       &                       & $6.27 \times 10^6$ & -0.383 & $-$ \\
\hline
1c & \multirow{4}{*}{0.10} & \multirow{4}{*}{0.05} & $3.30 \times 10^7$ & -0.481 & 0.0981 \\           
2c &                       &                       & $2.44 \times 10^7$ & -0.383 & 0.0917 \\
3c &                       &                       & $1.68 \times 10^7$ & -0.292 & 0.0844 \\
4c &                       &                       & $9.40 \times 10^6$ & -0.207 & $-$

\end{tabular}
\end{center}
\end{minipage}
\end{table}

Fig. \ref{fig:Evolution} shows results of the numerical simulations for different Doppler widths $\lambda^{}_{\Theta} = 0.20, 0.10$, and $0.05$ where the initial spectral widths are $\sigma_{\rm init} = 0.2$ for the left panel and $\sigma_{\rm init} = 0.1$ for the right panel, respectively.
Black lines in Fig. \ref{fig:Evolution} show the initial spectra $N_0(x)$, and red and blue lines are photon spectra $N(x, y)$ at $y = 0.2 y_0$ and at $y = 0.4 y_0$, respectively.
We find that solitary structures are formed intermittently and shift to lower frequency, i.e., direction of their motion is basically determined by the left-hand side of Eq. (\ref{eq:ExtendedKompaneetsEquation}) and their solitary form results from the first term of the right-hand side of Eq. (\ref{eq:ExtendedKompaneetsEquation}).
The heights of each solitary structure increase with time without changing their logarithmic width.
The smaller value of $\lambda^{}_{\Theta}$, i.e., the colder plasma, the more and the narrower solitary structures are formed.
This behavior is consistent with what is inferred from the steady-state solution (Eq. (\ref{eq:SteadyStateSolution})).
On the other hand, numerical simulations without the third-derivative in Eq. (\ref{eq:ExtendedKompaneetsEquation}) gives an unstable shock-like discontinuous structure rather than solitary structures with the use of the same scheme.

To characterize the solitary structures, we consider a zeroth-order log-normal distribution for each solitary structure and fit some of the results at $y = 0.4 y_0$ with a function
\begin{eqnarray}\label{eq:Fitting}
	F(x)
	& = &
	N_0(x)
	+
	\sum_i 
	n_i
	\exp\left( - \frac{(\ln x  - m_i)^2}{2 w^2} \right),
\end{eqnarray}
where we adopt a half width of $w = \lambda^{}_{\Theta} / 4$ considering Eq. (\ref{eq:SteadyStateSolution}).
Fitted values of $n_i$, $m_i$ and the separation between the neighboring peaks $\Delta m_i$ (for example, $\Delta m_{\rm 1a} = m_{\rm 2a} - m_{\rm 1a}$) are listed in Table \ref{tbl:Fitting}.
Numbering of solitary structures is found on the insets in Fig. \ref{fig:Evolution}; $i = {\rm 1a - 4a}$ for $(\sigma_{\rm init}, \lambda^{}_{\Theta}) = (0.20, 0.10)$, $i = {\rm 1b - 7b}$ for $= (0.20, 0.05)$ and $i = {\rm 1c - 4c}$ for $= (0.10, 0.05)$, respectively.
We do not fit all the structures because some structures have a smaller amplitude than $N_0(x)$.
For example, we see more structures on the right of the structure numbered `4c' of the inset on the right panel in Fig. \ref{fig:Evolution}.
Logarithmic widths of the structures are well characterized by the Doppler width $\lambda^{}_{\Theta}$.
However, the spacing of neighboring peaks $\Delta m_i$ does not follow Eq. (\ref{eq:SteadyStateSolution}), i.e., $\Delta m_i \ne \lambda^{}_{\Theta}$ (see Table \ref{tbl:Fitting}), or rather $\Delta m_i$ is larger for lower frequency peaks.
This relates with larger $n_i$ at lower frequency, because the velocity of a wave is proportional to $N(x)$ in Eq. (\ref{eq:ExtendedKompaneetsEquation}).
Fitted values $m_{\rm 1a} < m_{\rm 1b}$ indicate that the velocity of solitary structures is slower for larger value of $\lambda^{}_{\Theta}$.
Although the velocity relates with energy loss of photons, it is not easy to discuss the dependence on $(\sigma_{\rm init}, \lambda_{\Theta})$ with Fig. \ref{fig:Evolution}, because number and height of structures are also different for different sets of $(\sigma_{\rm init}, \lambda_{\Theta})$ (see the discussion about photon energy transfer in Section \ref{sec:Discussion}).

\subsection{Discussion}\label{sec:Discussion}

Radio emissions from pulsars (e.g. Ref. \cite{Hankins&Eilek07}) and also fast radio bursts (e.g. Refs. \cite{Thornton+13}) have extremely large brightness temperature.
Although the optical depth to ICS $\tau_{\rm ind}$ would attain to unity for these objects (c.f. Refs. \cite{Lyubarsky08, Tanaka&Takahara13, Wilson&Rees78}), the past studies did not discuss what kind of signatures are expected to be imprinted in their observed spectra.
Our present study predicts a spectral break at the frequency $\nu_{\rm ind}$ corresponding to $\tau_{\rm ind}(\nu_{\rm ind}) \gtrsim 0.2$ and solitary structures below $\nu_{\rm ind}$ in the photon spectra (red lines in Fig. \ref{fig:Evolution}).
In this regard, the discrete emission bands in the dynamic spectra of the giant radio pulse occurred at the interpulse phase in the Crab pulsar reported by Hankins \& Eilek \cite{Hankins&Eilek07} (`zebra bands') are intriguing phenomena.
If we interpret their reported value $\Delta \ln \nu = \Delta \nu / \nu \sim 0.06$ by ICS, the value is not far from $\Delta m_i$ for $\lambda^{}_{\Theta} = 0.05$ in Table \ref{tbl:Fitting} corresponding to the electron temperature of $\sim$ a few $\times 10$ eV.
However, for applications to realistic astrophysical situations, we should take account for anisotropic photon distributions and for relativistic effects of both bulk and thermal motions of plasmas.
These effects will be studied in subsequent papers.

\begin{figure}[t]
\begin{center}
\includegraphics[scale=0.70]{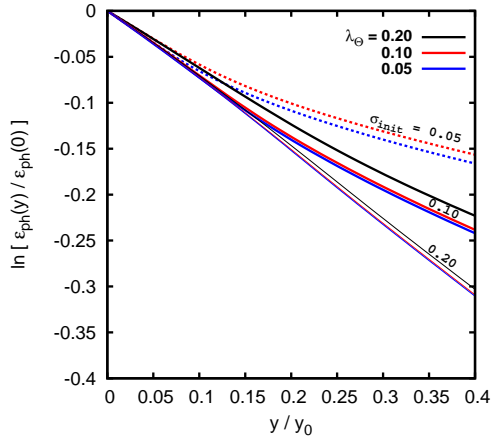}
\end{center}
\caption{
	Evolution of the normalized photon energy density $\epsilon_{\rm ph}(y) / \epsilon_{\rm ph}(0)$ with the normalized time $y / y_0$.
	The vertical axis is the natural logarithm of $\epsilon_{\rm ph}(y)/\epsilon_{\rm ph}(0)$.
	Thick and thin lines correspond to $\sigma_{\rm init} = 0.20$ (left panel in Fig. \ref{fig:Evolution}) and $= 0.10$ (right panel in Fig. \ref{fig:Evolution}).
	We also show the results for $\sigma_{\rm init} = 0.05$ (dotted lines) for two different Doppler widths of $\lambda^{}_{\Theta} = 0.10$ (red) and $0.05$ (blue), respectively.
}
\label{fig:EnergyDensityEvolution}
\end{figure}

While the quantities $\int N(x, y) dx$ and $\int \ln N(x, y) dx$ are conserved with time, the photon energy density $\epsilon_{\rm ph}(y) = \int x N(x, y) d x$ decreases with time, i.e., the energy of photons is transfered to plasmas by ICS \cite[][]{Levich&Sunyaev70, Blandford73}.
The rate for the energy transfer by ICS is estimated from the integration of Eq. (\ref{eq:InducedOnly}), $d \epsilon_{\rm ph}(y)/ d y = - \int N^2(x, y) dx$ (see Eq. (\ref{eq:PhotonEnergyEvolution})).
Considering the initial phase of evolution $y \ll y_0$, i.e., $N(x, y) \approx N_0(x)$, we estimate the rate as
\begin{eqnarray}\label{eq:InitialPhotonEnergyTransfer}
	\frac{d \epsilon_{\rm ph}(y)}{d y}
	& \approx &
	- 2 \sigma_{\rm init} N^2_{\rm init}, \nonumber \\
	& \approx &
	- \frac{\epsilon_{\rm ph}(y)}{y_0} ~\mbox{for $y \ll y_0$.}
\end{eqnarray}
where we approximate $\epsilon_{\rm ph}(y) = \int x N(x, y) d x \approx 2 x_{\rm init} \sigma_{\rm init} N_{\rm init}$ and recall $y_0 = x_{\rm init} / N_{\rm init}$ for the last expression. 
Eq. (\ref{eq:InitialPhotonEnergyTransfer}) can be applicable to describe an initial evolutionary phase and gives exponential loss of the photon energy with time $\epsilon_{\rm ph}(y) \approx \epsilon_{\rm ph}(0) \exp(-y/y_0)$.

Fig. \ref{fig:EnergyDensityEvolution} plots evolution of the normalized photon energy density with time for the results of Fig. \ref{fig:Evolution}.
We also show the case for an initial width of $\sigma_{\rm init} = 0.05$ as dotted lines in Fig. \ref{fig:EnergyDensityEvolution}.
Thin lines ($\sigma_{\rm init} = 0.20$) show the exponential energy loss, although the slope is not exactly $\ln [\epsilon_{\rm ph}(y) / \epsilon_{\rm ph}(0)] = -y/y_0$.
Thick ($\sigma_{\rm init} = 0.10$) and dotted ($\sigma_{\rm init} = 0.05$) lines deviate from the exponential-law (thin lines) at $y \sim 0.15 y_0$ and $y \sim 0.05 y_0$, respectively.
Dotted red line $(\sigma_{\rm init}, \lambda_{\Theta}) = (0.05, 0.10)$ and thick black line $(\sigma_{\rm init}, \lambda_{\Theta}) = (0.10, 0.20)$ show a bit different slope even at $y \lesssim 0.05 y_0$ from other lines and these exceptional behaviors are characterized by a relatively large Doppler wavelength $\lambda_{\Theta} = 2 \sigma_{\rm init}$, which will be discussed later in this section. 
The energy transfer from photons to plasmas by ICS is effective for the case of broad initial spectra than narrow ones (the blue lines are always below the red lines for the same $\sigma_{\rm init}$).
This means that the deviation of $N(x, y)$ from the initial spectrum $N_0(x)$ is faster for narrower spectra and this behavior is found from the inset on the right panel of Fig. \ref{fig:Evolution} (the blue line is well below the green line compared with those on the left panel of Fig. \ref{fig:Evolution}).
Although we see a temperature dependence of the energy transfer for the same initial width, Eq. (\ref{eq:InitialPhotonEnergyTransfer}) has no information of electron temperature explicitly.
The third-derivative in Eq. (\ref{eq:ExtendedKompaneetsEquation}) may play as a heating term because larger $\lambda_{\Theta}$ (higher temperature) gives slower energy transfer in Fig. \ref{fig:EnergyDensityEvolution}.
This is consistent with the discussion at the steady-state solution Eq. (\ref{eq:SteadyStateSolution}), i.e., the photon flux is written as $\approx B^2 - A^2$ for $\Theta \ll 1$ and a contribution from $\Delta I_{(k p^2)}$ ($A$) resists that from $\Delta I_{(k)}$ ($B$).

We should note the case of $\lambda_{\Theta} = 2 \sigma_{\rm init}$, i.e., full width of the initial spectrum is almost the same as the Doppler width.
For $\lambda_{\Theta} > 2 \sigma_{\rm init}$, evolution of photon spectra is unstable at least in our numerical scheme.
In those calculations, although all of the results shown in this paper have no time variation of $N(x > x_{\rm init}, y)$ (see Fig. \ref{fig:Evolution}), there appears unstable features at $x \gtrsim x_{\rm init}$ for $\lambda_{\Theta} > 2 \sigma_{\rm init}$.
This unstable feature is not improved by higher numerical resolutions of both $x$ and $y$.
Currently, it is uncertain whether the exceptional behaviors of dotted red line $(\sigma_{\rm init}, \lambda_{\Theta}) = (0.05, 0.10)$ and thick black line $(\sigma_{\rm init}, \lambda_{\Theta}) = (0.10, 0.20)$ in Fig. \ref{fig:EnergyDensityEvolution} is numerical or physical.
However, we should also take care of our formulation (Eq. (\ref{eq:BoltzmannEquation})) for a narrow spectrum $\sigma_{\rm init} \ll \lambda_{\Theta}$.
Galeev and Syunyaev (1973) \cite{Galeev&Syunyaev73} argued that, when $\sigma_{\rm init} \ll \lambda_{\Theta}$, collective behaviors of plasmas are important rather than the interaction with free electrons, i.e., Compton scattering.
This is the reason why we do not study the case for $\lambda^{}_{\Theta} > 2 \sigma_{\rm init}$ more in this paper, although study of this regime is important for the application to laboratory experiments.

We should also note that whole of present paper is based on the assumption $\Theta =$ const.
As seen in Fig. \ref{fig:EnergyDensityEvolution}, photons clearly lose energy by ICS and transfer their energy to plasmas.
The time-scale of energy transfer to electrons is estimated from Eq. (\ref{eq:InitialPhotonEnergyTransfer}) as
\begin{eqnarray}\label{eq:HeatingRate}
	\left( \frac{1}{\epsilon_{\rm pl}(y)} \frac{d \epsilon_{\rm pl}(y)}{d y} \right)^{-1}
	& = &
	\left( - \frac{1}{\epsilon_{\rm pl}(y)} \frac{d \epsilon_{\rm ph}(y)}{d y} \right)^{-1} \nonumber \\
	& \approx &
	y_0 \frac{\epsilon_{\rm pl}(y)}{\epsilon_{\rm ph}(y)} ~\mbox{for $y \ll y_0$,}
\end{eqnarray}
where $\epsilon_{\rm pl} \approx n_{\rm e} \Theta$ is the energy density of a plasma divided by $m_{\rm e} c^2$.
We need to take into account evolution of a plasma temperature for evolution beyond this time-scale (c.f. Refs. \cite{Reinish76a, Reinish76b}).
Note that Eq. (\ref{eq:HeatingRate}) is allowed to be used even for $\epsilon_{\rm pl}(y) > \epsilon_{\rm ph}(y)$ (c.f. Ref. \cite{Zel'dovich75}).
Our calculation needs only $x_{\rm init}$, $N_{\rm init}$, $\sigma_{\rm init}$ and $\lambda_{\Theta}$ that determine $y_0$, $\epsilon_{\rm ph}(0)$, and $\Theta$, i.e., we do not specify $n_{\rm e}$ and $t$ separately.
Typical time-scales of photon cooling $t_{\rm ph}$ (Eq. (\ref{eq:InitialPhotonEnergyTransfer}) and also Eq. (\ref{eq:ExtendedKompaneetsEquation})) is found from $y_{\rm ph} \sim y_0$ and that of electron heating $t_{\rm pl}$ (Eq. (\ref{eq:HeatingRate})) is found from $y_{\rm pl} \sim y_0 (\epsilon_{\rm pl} / \epsilon_{\rm ph})$.
These are expressed as
\begin{eqnarray}\label{eq:TimeScales}
	t_{\rm ph} 
	& \sim &
	\frac{x_{\rm init}}{N_{\rm init} \Theta}
	\frac{1}{n_{\rm e} \sigma^{}_{\rm T} c}, ~~ 
	t_{\rm pl} 
	\sim
	\frac{1}{2 \sigma_{\rm init} N^2_{\rm init} \Theta^4}
	\frac{\pi^2 \lambdabar^3_{\rm e}}{\sigma^{}_{\rm T} c}, 
\end{eqnarray}
where the energy densities of photons and electrons are $\epsilon^{\ast}_{\rm ph} = \hbar c \Theta^4 / (\pi^2 \lambdabar^4_{\rm e}) \int x N(x) d x$ and $\epsilon^{\ast}_{\rm pl} \sim n_{\rm e} \Theta m_{\rm e} c^2$ including dimension, respectively.
For given $x_{\rm init}$, $N_{\rm init}$, $\sigma_{\rm init}$ and $\lambda_{\Theta}$, photons lose energy before heating electrons ($t_{\rm ph} \ll t_{\rm pl}$) for sufficiently large $n_{\rm e}$.
For example, we obtain $t_{\rm ph} \lesssim t_{\rm pl} \sim$ 10 ps for $n_{\rm e} \gtrsim 10^{22} {\rm cm^{-3}}$ (close to conditions of laser experiments) with the parameters which we used in this paper.

We finally discuss differences from past calculations of spectral evolution by ICS \cite{Montes79, Coppi+93}.
Fig. 1 of Coppi et al. (1993) \cite{Coppi+93} is spectral evolution for isotropic case.
Basically, they solved Eq. (\ref{eq:InducedOnly}) but including numerical viscosity, i.e., their calculations did not have information of a plasma temperature $(\lambda_{\Theta})$.
Although their result (the right panel of Fig. 1 of Coppi et al. (1993) \cite{Coppi+93}) showed two solitary structures that have the same logarithmic width but their behaviors seem different from ours, such as the height of the low frequency structure is smaller than that of the high frequency one.
We consider that these structures are numerical, i.e., combination of the numerical viscosity and logarithmically-spaced frequency grids in their calculations.
Montes (1979) \cite{Montes79} solved the integro-differential equation given by Zel'dovich et al. (1972) \cite{Zel'dovich+72}, but with some simplifications.
Fig. 2 of Montes (1979) \cite{Montes79} (the calculation for an initially broad Gaussian spectrum) shows solitary structures of the linearly same width.
We consider that Eq. (27) of Montes (1979) \cite{Montes79} is over simplified, especially, their assumption $\Delta \nu \approx \Theta^{1/2} \nu_0 =$ const. in their Eq. (20) is crucial, while we consider the case $\Delta \nu \approx \Theta^{1/2} \nu$ \cite[][]{Zel'dovich&Sunyaev72}.

\section{Conclusions}\label{sec:Conclusions}

In this paper, we study evolution of photon spectra when ICS dominates.
To get rid of the well-known difficulty of the nonlinear convection equation (Eq. (\ref{eq:InducedOnly})), we consider the higher-order Kompaneets equation.
We obtain the new equation (Eq. (\ref{eq:ExtendedKompaneetsEquation})) that describes evolution of photon spectrum by ICS and that overcomes the difficulty.
The second-order induced term $\Delta I_{(k p^2)}$ (Eq. (\ref{eq:Order_kp^2_Induced})) obtained from the higher-order Kompaneets equation improves the formulation.
In addition, Eq. (\ref{eq:ExtendedKompaneetsEquation}) has the steady-state analytic solution (Eq. (\ref{eq:SteadyStateSolution}) and Fig. \ref{fig:SteadyStateSolution}), which is linearly stable.
The steady-state analytic solution predicts the formation of solitary structures of logarithmically same width in frequency-space and the $\Delta I_{(k p^2)}$ term plays as a heating term against the ICS term $\Delta I_{(k)}$, which is the pure cooling term.

We also study evolution of photon spectra by ICS numerically.
ICS intermittently forms solitary structures moving toward lower frequency (Fig. \ref{fig:Evolution}) and these behaviors are consistent with predictions by some of past studies \cite[][]{Zel'dovich+72, Zel'dovich&Sunyaev72}.
The solitary structures have the logarithmically same width well characterized by the Doppler width $\lambda_{\Theta}$ and this behavior is also inferred from the steady-state solution.
On the other hand, the spacing between peak frequencies of the structures does not follow Eq. (\ref{eq:ExtendedKompaneetsEquation}) $(\Delta m_i \ne \lambda_{\Theta})$.
The number and height of solitary structures depend on $\lambda_{\Theta}$.

The results of our numerical simulation satisfy the two invariants of motion, conservations of photon number $\int N(x) dx$ and of the quantity $\int \ln N(x) dx$.
On the other hand, the energy density of photons $\int x N(x) dx$ is transferred to electrons (Fig. \ref{fig:EnergyDensityEvolution}).
Energy density decays exponentially in an initial phase and this behavior is almost consistent with the analytic estimate (Eq. (\ref{eq:InitialPhotonEnergyTransfer})).
The energy transfer from photons to plasmas by ICS is effective for the case of broad initial spectra such as expected in astrophysical situations.
The numerical results shown in the present paper are only initial phases of evolution $y \le 0.4 y_0$.
We need more optimized numerical scheme to study evolution for $y > 0.4 y_0$.

\section*{Acknowledgments}

S.J.T. would like to thank Y. Ohira, R. Yamazaki, Y. Sakawa and F. Takahara for useful discussions.
We would also like to thank the anonymous referee for his/her helpful comments.
This work is supported by JSPS Research Fellowships for Young Scientists (S.T., 2510447).

\end{document}